\documentclass[11pt]{article} 

\usepackage{amsfonts}
\usepackage{amsbsy} 

\advance \topmargin by -\headheight
\advance \topmargin by -\headsep
\evensidemargin \oddsidemargin
\advance\hoffset by -3mm  
\advance\voffset by  8mm  

\begin{document}
\newcommand{\hk}{hyperK\"{a}hler}
\newcommand{\ka}{K\"{a}hler}
\newcommand{\sas}{Sasakian}
\newcommand{\tris}{tri--Sasakian}
\newcommand{\ben}{\begin{equation}}
\newcommand{\een}{\end{equation}}
\newcommand{\ti}[1]{\ensuremath{\tilde{#1}}}
\newcommand{\bs}{\ensuremath{\boldsymbol{\omega}}}
\newcommand{\re}{\ensuremath{\mathbb{R}}}
\newcommand{\q}{\ensuremath{\mathbb{H}}}
\newcommand{\euc}{\ensuremath{\mathbb{E}}}
%
\title{Cones,  tri-Sasakian structures and superconformal invariance}
\author{G.W. Gibbons\thanks{e-mail:gwg1@damtp.cam.ac.uk}\ \  and \    
        P. Rychenkova\thanks{e-mail:pr201@damtp.cam.ac.uk} \\
        DAMTP, University of Cambridge, Silver Street \\ 
        Cambridge, CB3 9EW, U.K.} 
\date{\today}
\maketitle
\bigskip
\centerline{DAMTP-1998-129}
\begin{abstract}
In this note we show that rigid $N=2$ superconformal hypermultiplets
must have target manifolds which are cones over tri-Sasakian
metrics. We comment on the relation of this work to cone-branes and the
AdS/CFT correspondence.
\end{abstract}
\section{Introduction}
There has recently been great interest in rigid conformally invariant
supersymmetric field theories. In particular de~Wit, Kleijn and
Vandoren \cite{deWit} have studied $N=2$ models containing
hypermultiplets taking values in a \hk\ manifold $({\cal M}, g_{\mu\nu},
I^{\ \mu}_{a\ \ \nu})$, where $\mu,\nu = 1,\ldots,4k = dim{\cal M}$ and $a =
1,2,3$. They find the following necessary condition that the target manifold
admits an infinitesimal dilatation invariance:
$({\cal M},g_{\mu\nu})$ admits a vector field $X^\mu$ such that
\begin{equation} \label{dilatation}
\begin{tabular}{|c|} \hline 
$ X^\mu_{\ ;\nu} = \delta^\mu_{\ \nu} $ \\ \hline
\end{tabular}
\end{equation}
In this note we point out that condition (\ref{dilatation}) implies
(regardless of any \hk\ condition) that $\cal M$ is a cone, $C(B)$, over
a base manifold $B$, i.e. in coordinates $x^\mu = (r, x^i)$, $i =
1,\ldots,dim{\cal M} - 1$, the metric $g_{\mu\nu}$ is
\ben \label{conemet}
g_{\mu\nu}\,dx^\mu\,dx^\nu = dr^2 + r^2\,h_{ij}\,dx^i\,dx^j\ , 
\een
where $h_{ij}$ is the metric on the base $B$ which depends only on
$x^i$.  Moreover, in these
coordinates
\[
X = r\frac{\partial}{\partial r}
\]
and so the dilatation acts as
\[
(r, x^i)\  \rightarrow\ (\lambda r, x^i),\ \ \ \ \lambda\in \re^+\ .
\]
The differential operator $X = r\partial/\partial r$ is sometimes
called the Eulerian vector field.

In the case that $({\cal M}, g_{\mu\nu})$ is a Ricci-flat \ka,
condition (\ref{dilatation}) implies that the vector field
$X^\mu$ is holomorphic, that the base manifold $B$ carries a \sas\
structure and hence the metric $g_{\mu\nu}$ admits a holomorphic
Killing vector field
\[
Y^\mu = I^\mu_{\ \nu}\,X^\nu\ , 
\]
where $I^\mu_{\ \nu}$ is the complex structure. Presumably this case
arises in $N=1$ rigid superconformally invariant theories
\cite{wittenklebanov}.  If $({\cal M}, g_{\mu\nu})$ is Ricci-flat then
$B$ must be Einstein--\sas. We do not know whether rigid $N=1$
superconformal invariance implies that the metric should be
Ricci-flat.

In the case that  $({\cal M}, g_{\mu\nu})$ is \hk, the base manifold
$B$ admits a \tris\ structure and the metric $g_{\mu\nu}$ also admits an
$SU(2)$ action by isometries which permutes the complex structures
$I_1, I_2$ and $I_3$. In this case the metric is necessarily
Ricci-flat and the base manifold is necessarily Einstein.

The organisation of the article is as follows. In section \ref{sect1}
we study equation (\ref{dilatation}) in an arbitrary metric
$g_{\mu\nu}$ and show that it leads to equation (\ref{conemet}). In
section \ref{sect2} we assume $({\cal M},g)$ is \ka. In section
\ref{sect3} we assume $({\cal M},g)$ is \hk. In section \ref{sect4} we
discuss a general homothety which does not satisfy
(\ref{dilatation}). Section \ref{sect5} contains examples and in section
\ref{sect6} we discuss applications of the results. We find
it remarkable how simply our main results follow from equation
(\ref{dilatation}) and, although there are a number of discussions of
cone geometries in the pure mathematics literature, we believe that
our simple and direct treatment will be especially appealing to field
theorists.  

\section{Cones and Dilatations} \label{sect1}

A manifold $({\cal M}, g_{\mu\nu})$ regardless of signature of
$g_{\mu\nu}$ admits a
conformal Killing vector field $X$ if and only if
\ben \label{isometry}
{\cal L}_X g_{\mu\nu} = \phi g_{\mu\nu} = X_{\mu ;\nu} + X_{\nu ;\mu}
\een
for some smooth function $\phi$. If $\phi$ is constant $X^\mu$ is said
to generate a {\em homothety}. If $X^\mu$ is hypersurface orthogonal, i.e.
\ben \label{hso}
X_\mu = \partial_\mu f \ \Longleftrightarrow \ X_{\mu ;\nu} = X_{\nu ;\mu}
\een
for some function $f$, we say that $({\cal M},g)$ admits as
infinitesimal {\em dilatation}. Since equations (\ref{isometry}) and
(\ref{hso}) are equivalent to equation (\ref{dilatation}) this is the
situation we are interested in. It follows that
\ben \label{covariantf}
\nabla_\mu\,\nabla_\nu f = g_{\mu\nu}.
\een 
Moreover, defining
\[
V = g^{\mu\nu}\partial_\mu f\,\partial_\nu f = g_{\mu\nu} X^\mu X^\nu\
,
\]
we have
\ben \label{fVeqn}
\partial_\mu\, V = 2\partial_\mu\, f\ .
\een 
We pick the arbitrary constant of integration such that
\[
V = 2f\ .
\]
Now we pick $f$ as one of the coordinates and find the metric to be
\ben \label{fmet}
ds^2 = \frac{df^2}{2f} + g_{ij}(x^r,f)\,dx^i\,dx^j\ .
\een
There is no cross term in the metric because $X^\mu$ is orthogonal to
the surface $f = const$.

Finally, we write out the equation
\ben \label{iso}
{\cal L}_X g_{\mu\nu} = g_{\mu\nu ,\lambda} X^\lambda +
g_{\mu\lambda} X^\lambda_{\ ,\nu} + g_{\nu\lambda} X^\lambda_{\ ,\mu}
= 2g_{\mu\nu}\ .
\een
Using the fact that
\[
X^\mu = g^{\mu\nu} \partial_\nu f = 2 f \delta^\mu_{\ 0}\ ,
\]
and substituting (\ref{fmet}) into (\ref{iso}), we obtain for the
$(i,j)$ component 
\ben \label{partialeqn}
f\ \frac{\partial g_{ij}}{\partial f} = g_{ij}\ .
\een
If we define 
\[
r^2 = 2f\ ,
\]
the solution of (\ref{partialeqn}) may be written as
\[
g_{ij} = r^2 h_{ij}(x^k)\ ,
\]
and the basic result (\ref{conemet}) follows.

In fact we need not assume that the metric $g_{\mu\nu}$ is Riemannian,
but if $X^\mu$ were time-like we would need to adjust the signs in
(\ref{conemet}). Note that from (\ref{fVeqn}) we cannot have $X^\mu$
light-like. From (\ref{covariantf}) we find that
\[
\nabla_\lambda\,\nabla_\mu\,\nabla_\nu f = 0\ . 
\]
But $(\nabla_\mu\,\nabla_\nu-\nabla_\nu\,\nabla_\mu)\,X^\alpha =
R^\alpha_{\ \beta\mu\nu}\, X^\beta$ gives
\[
R^\alpha_{\ \beta\mu\nu}\, X^\beta = 0\ ,
\]
which contracted on $\alpha$ and $\mu$ gives
\ben \label{ricci}
X^\beta\, R_{\beta\nu} = 0\ .
\een
Obviously (\ref{ricci}) is incompatible with $g_{\mu\nu}$ being an
Einstein metric with non-vanishing scalar curvature. However, it is
not incompatible with $g_{\mu\nu}$ being Ricci-flat, and indeed this
will be true if the metric $h_{ij}$ on the base is Einstein such that
\[
R_{ij} = (n-1)\,h_{ij}\ ,
\]
where $n = dim{\cal M}$. Such a metric $g_{\mu\nu}$ is called a
Ricci-flat cone.

Note that the assumption (\ref{hso}) that the homothety $X^\mu$ is hypersurface
orthogonal  played an essential role. For example Chave,
Tod and Valent \cite{tod} have exhibited a Ricci-flat (\hk)
four-metric which admits a homothety which is {\em not} hypersurface
orthogonal.

We conclude this section by remarking that cones have arisen in
supergravity theories under the guise of ``generalized dimensional
reduction.'' For example Pope {\em et al} \cite{pope} have used the scaling
invariance of supergravity theories to construct solutions which are
eleven-dimensional cones over ten-dimensional base manifolds.

\section{K\"{a}hlerian cones and \sas\ structures} \label{sect2}

Now we suppose that ${\cal M} = C(B)$ is a \ka\ manifold with a
(covariantly constant) complex structure $I$. We have
\begin{eqnarray}
{\cal L}_X I^\mu_{\ \nu} & =  & I^\mu_{\ \nu ;\sigma}X^\sigma -
I^\sigma_{\ \nu} X^\nu_{\ ;\sigma} + I^\mu_{\ \sigma} X^\sigma_{\ ;\nu} \\ 
& = & -I^\mu_{\ \nu} + I^\mu_{\ \nu} = 0 \nonumber\ . 
\end{eqnarray}
Hence necessarily $X$ is a holomorphic vector field. The reader is
cautioned however that since 
\[
{\cal L}_X \omega = d(\iota_X \omega) + \iota_X(d\omega) = 2\omega\ ,
\]
where $\omega$ is the \ka\ form with components $\omega_{\mu\nu} =
g_{\mu\sigma} I^\sigma_{\ \nu}$, we have 
\[
d(\iota_X \omega) = 2\omega
\]
and hence $X$ is not Hamiltonian and there is no conventional moment
map. 

The vector field
\ben \label{killing}
K^\mu = I^\mu_{\ \nu} X^\nu
\een
satisfies
\[
K_{\mu ;\nu} = \omega_{\mu\nu}\ .
\]
Thus $K$ is a Killing field and it is easily seen to be
holomorphic and to commute with $X$. In addition, $K$ is a Hamiltonian
vector field whose moment map is $f$ and hence
the level sets of the moment map coincide with the base manifold $B$.
$K^\mu$ is tangent to the base manifold $B$ and is therefore a
Killing field of the metric $h_{ij}$. 
From (\ref{killing}) we have 
\[
K_\mu K^\mu = X_\nu X^\nu = V\ .
\]
Thus the length of the vector $K^\mu$ is constant along base manifold
$B$. Choosing $V = 1$ as our base manifold we have the following
structure on $B$:
\begin{itemize}
\item a one-form $\eta_i = K_i$
\item a vector field $\xi^i = K^i$
\item an endomorphism $I^i_{\, j}$ of the tangent bundle $T(B)$ of $B$ 
\item a metric $h_{ij}$\ .
\end{itemize}
It is straightforward to check that $(B,\ \xi^i,\ \eta_i,\  I^i_{\
  j})$ satisfies the conditions for  a \sas\ manifold \cite{blair}.

Note that we have not assumed that $M$ is Ricci-flat. If we did so,
then $(B,h_{ij})$ would necessarily be an Einstein--\sas\ manifold.

\section{Hyper\ka ian cones and \tris\ structures} \label{sect3}
Now we suppose ${\cal M} = C(B)$ is \hk\  and hence necessarily
Ricci-flat. The base metric must therefore be Einstein. The vector
field $X$ is tri-holomorphic, i.e. it preserves the three complex
structures $I_a$ and their algebra.
There are three Killing vector fields tangent to $B$ and
commuting with $X$:
\[
K_a^\mu = I_{a\ \ \nu}^{\ \mu} X^\nu\ .
\]
However now
\[
{\cal L}_{K_a} I_b = -\, 2\, \epsilon_{abc}\  I_c
\]
and
\[
[K_a,K_b] = -\, 2\, \epsilon_{abc}\  K_c\ .
\]
Thus we have a  non-triholomorphic $SU(2)$ action on $\cal M$ which
again descends to the base manifold $B$. In fact we now have a \tris\
structure on $B$. Each generator $K_a$ of the $SU(2)$ action is
holomorphic with respect to its own complex structure $I_a$ and $f$ is the
associated moment map. The emergence of an extra $SU(2)$ isometry group
was noticed in \cite{deWit}. For more information and references to the
mathematical literature on \tris\ structures the reader is directed to
\cite{hull}.

\section{Hypersurface non-orthogonality} \label{sect4}

As we emphasised above the assumption (\ref{hso}) that the homothetic
Killing field $X$ is hypersurface orthogonal is essential for our
result. Suppose that $X$ is a homothety which is not hypersurface
orthogonal. Defining
\[
F_{\mu\nu} = \partial_\mu X_\nu - \partial_\nu X_\nu\ ,
\]
one finds that 
\[
{\cal L}_X\  I^\mu_{\ \nu} = I^\mu_{\ \sigma} F^\sigma_{\ \nu} -
F^\mu_{\ \sigma} I^\sigma_{\ \nu}\ .
\]
Thus, in general, $X$ need not be holomorphic with respect to any
complex structure. Moreover, defining $K$ as in (\ref{killing}) we
have
\[
{\cal L}_K g_{\mu\nu} = \omega_{\mu\sigma} F^\sigma_{\ \nu} +
\omega_{\nu\sigma} F^\sigma_{\ \mu}\ . 
\]
Therefore we do not necessarily have an extra isometry. One might
wonder whether, assuming $\cal M$ is \hk, any non-trivial homothety
could exist. In their paper  Chave {\em et al} \cite{tod} have given a
family of four-dimensional \hk\ metrics with a tri-holomorphic
homothety which is not hypersurface orthogonal. The metric is of the
form
\[
e^{2t}\left\{\frac{1}{W} (dt + A)^2 + W \ti{h}_{ij} dx^i dx^j\right\}\ ,
\]
where the metric in the base gives a $(4,0)$ sigma model and $W$ and
$A$ satisfy monopole-like equations.

\section{Symmetry enhancement and examples} \label{sect5}

There is no shortage of examples of \tris\ manifolds (see
\cite{hull} and references therein.) However, unless we take $B$ to be
a sphere $S^{4k-1}$ with its standard \tris\ structure, the manifold
$\cal M$ will be singular at
the vertex $r=0$. In some cases the singularity may be removed to
give a non-singular \hk\ manifold which no longer admits an exact
dilatation symmetry but continues to do so approximately at
infinity. 

The obvious example are the ALE cones for which the base $B$
is $S^3/\Gamma$, where $\Gamma$ is a finite subgroup $\Gamma \subset
SU(2) \subset SO(4)$. They may be thought of as the quotient of
$\re^4$ by $\Gamma$ with an orbifold fixed point at the origin. As is
well known \cite{kronheimer} this may be blown up to give a
non-singular manifold. It is instructive to consider the multi-centre
case (see e.g. \cite{cyclicale}). This may be constructed \cite{us} as
the  \hk\ quotient
\[
\q^{m+1} / /\  (U(1))^m\ .
\]
The level sets of the moment maps are
\ben \label{alemomentmap}
\mu_\alpha = q_\alpha\, i\, \bar{q}_\alpha + q\, i\, \bar{q} =
\zeta_\alpha\ ,
\een
where $\alpha = 1,\ldots, m$ and the quaternions $(q_\alpha,q)$
parametrise $\q^{\, m+1}$. The quantities $(\zeta_\alpha - \zeta_\beta)$
correspond to the relative separation of the centres. Now let
$\zeta_\alpha \rightarrow 0$ for all $\alpha$. We get the orbifold
limit in which the size of all two-cycles shrinks to zero. In the same
limit  the level sets (\ref{alemomentmap}) become
invariant under the dilatation of $\q^{\, m+1}$ given by 
\[
(q_\alpha, q)\  \longrightarrow\  (\lambda q_\alpha, \lambda q)
\]
which descends to the quotient orbifold. Thus the appearance of
the dilatation symmetry is associated with the shrinking of two-cycles. 
Note that a general ALE metric has no $SU(2)$ isometry,
tri-holomorphic or not. As we approach the orbifold limit the isometry
group is enhanced to include $\re_{\, +}\times SU(2)$ where $\re_{\, +}$
corresponds to dilatations. Note also that although there are many
\hk\ manifolds with non-triholomorphic $SU(2)$ actions they are not
all cones. Neither are they necessarily asymptotically conical. For example
all BPS monopole moduli spaces admit such an $SU(2)$ or $SO(3)$ action
which arises from rotations in physical space, but they do not
admit dilatations because of the scale set by the monopole
mass. 

An interesting question for further study is whether one can construct
non-locally flat dilatation invariant \hk\ manifolds using the \hk\
quotient construction of a flat space.

\section{Discussion} \label{sect6}

Cones over \sas\ and \tris\ manifolds have recently made an appearance
in M-theory \cite{hull,wittenklebanov}. One considers $p$-brane
solutions of  the form
\[
H^{-\alpha}\, (-dt^2 + d{\bf x}_p^2) + H^{\frac{2}{\beta}}\, g_C
\]
with $H = 1+(\alpha/r)^\beta$ and $g_C$ the metric on a Ricci-flat
cone with base
$B$. These interpolate between $\euc^{\, p,1}\times C(B)$ at infinity and
$AdS_{p+2}\times B$ near the throat. This supergravity solution
corresponds to a large number, $k$, of Dirichlet $p$-branes.

The general belief is that the $U(1)$ factor of the world-volume $U(k)$ gauge
theory is associated with the centre of mass motion. The $(9-p)$ scalars
give the transverse coordinates of the branes. The
amount of supersymmetry of the world-volume theory is expected to agree with
amount of supersymmetry of the supergravity background. 

If $p=3$ it is tempting to make a
connection with the four-dimensional rigid $N=2$ conformally invariant
theories considered in \cite{deWit}. However although cones appear
both in the construction of the bulk space-time and as the target space
of the world-volume theory the cones are, in general, not the same. The
base $B$ of
the cone used to construct the bulk space-time is five-dimensional and
Einstein--\sas. The base of the cone of the target space of a
putative $N=2$ world-volume theory must be $(4n-1)$-dimensional and
\tris. Moreover, the amounts of supersymmetry of the supergravity
solution and
the world-volume theory do not agree.  

We have a better bet with $N=1$ superconformal theories based on
six-dimensional Calabi--Yau cones. The
idea would be that the six centre of mass coordinates of the
three-branes should assemble into
three complex Higgs fields of the world-volume theory. This appears to
coincide with the
example considered in \cite{wittenklebanov}: one takes $B
= (SU(2)\times SU(2))/U(1)$ with its Einstein--\sas\ structure. 

For the M2-brane the cone of the supergravity solution is
seven-dimensional and this could
be taken to be \tris. One might then contemplate identifying a
hypermultiplet of the $(2+1)$-dimensional world-volume theory with the
coordinates transverse to the M2-brane. However this looks rather
artificial and suggests that one should look elsewhere for the
geometrical origin of the hypermultiplets. By analogy with our
discussion for the D3-brane it would seem to be more fruitful to
follow \cite{wittenklebanov} and consider three-dimensional $N=2$
world-volume theories\footnote{The counting $N=2$ is from the
  three-dimensional point of view.} associated to
an eight-dimensional Calabi--Yau cone. The case analysed in
\cite{wittenklebanov}  is
$B = SO(5)/SO(3)$ with its standard Einstein-\sas\ structure.

\subsection*{Acknowledgements}
GWG would like to thank B.~de Wit and S.~Vandoren for raising the
question of the geometrical significance of the equation
(\ref{dilatation}) during the Ecole d'\`et\'e at ENS in August
1998. He would also like to thank Eugene Cremmer and other members of
the ENS for their hospitality.
PR thanks Trinity College, University of Cambridge, for a studentship.

Our discussion in section \ref{sect6} owes a great deal to the
comments of S.Vandoren and J.Figueroa--O'Farrill on an earlier draft
of this paper. We also thank
J.Figueroa--O'Farrill for drawing our attention to an earlier paper by
K.Galicki \cite{gal} who considered locally supersymmetric $N=2$ theories,
rather than rigid superconformal theories considered in \cite{deWit},
and related them to cones. However he did not explicitly derived
equation (\ref{dilatation}) and its consequences.

\end{document}